\documentclass[preprint,showpacs,preprintnumbers,amsmath,amssymb]{revtex4}

\usepackage{graphicx}
\usepackage{dcolumn}
\usepackage{bm}
\begin{document}

\title{Charge Transport in Non-Diluted Conjugated Polymers}
\author{Yohai Roichman and Nir Tessler}
\affiliation{Electrical Engineering Department, Technion, Haifa
32000, Israel}

\date{\today}

\begin{abstract}
We present a unified calculation method for variable range hopping
transport with a varying charge concentration and a varying
applied electrical field. We demonstrate that the major
differences between the transport properties measured at high
concentration and low concentration can be explained within this
framework. In particular the difference between the measured
mobility, and mobility activation energy in polymer field effect
transistors and polymer light emitting diodes is explained. A
theoretical method to extract the charge carrier density of state
from the transport measurements of non-diluted materials is
proposed.
\end{abstract}

\pacs{72.10.-d, 72.80.Le, 73.61.Ph}     
\maketitle

Charge transport in conjugated polymers is often described in
terms of thermal assisted tunneling (hopping) \cite{r1,r2} or
polaronic transfer mechanism \cite{r3} in a disordered array of
localized states \cite{r4}. As the density of states of the charge
carrier has not been directly measured for most of the conjugated
polymers, a Gaussian density of states is often assumed. The
Gaussian disorder model (GDM) \cite{r5} and related models with
energetic spatial correlations between the states \cite{r6} have
been solved previously for non-interacting carriers, or explicitly
for low charge density. Such calculations are typically successful
in describing the field and temperature dependence of the measured
mobility in the steady state regime. The dynamic transport
properties at the short time scales, where the charge carriers are
far from thermodynamic equilibrium, have been described using
different methods \cite{r7,r8}. However, all of the mentioned
calculations are based on the assumption that the time average
occupation of the sites is small and hence, the effect of the
charge concentration is negligible.

Two types of experiments demonstrated that the transport depends
on the charge density: the super linear increase of current with
increasing molecular doping, and the trans-conductance
measurements in polymeric field effect transistors (FETs)
\cite{r9}. Recently, we demonstrated a mobility enhancement in
poly-[2-methoxy-5-(2'-ethyl-hexiloxy)-p-phenylenevinylene]
(MEH-PPV) FET by a factor  of $\sim$10 due to charge concentration
build up \cite{r10} reaching values of up to $3\cdot10^{-5}$
cm$^2$/V sec. In light emitting diodes (LEDs) made from the same
polymer the mobility depends strongly on the applied field, with a
zero field mobility value of $1\cdot10^{-6}$ cm$^2$/V sec
\cite{r11} (see also \cite{r12}). Typically, FETs operate at much
lower electrical fields compared to LEDs and the charge residence
time in the device is much longer. Therefore, the increase of the
mobility in FETs cannot be explained in terms of electric field
dependent mobility or by the dynamical transport behavior at short
time scales.

Recently, the mobility \cite{r13} and the diffusion coefficient
\cite{r14} dependence on charge concentration has been calculated
using a low field approximation of the GDM, successfully
describing the molecular doping effect. Using the energy
correlation model with traps, the mobility dependence on
electrical field at different charge concentration was calculated
\cite{r15}, resulting in a similar field dependence of the
mobility for different carrier densities. In this letter we
suggest a unified calculation of the concentration and electrical
field dependence, using a semi-classical mean medium approximation
(MMA) \cite{r16} and demonstrate that the fundamental behavior of
the different devices is recovered. Moreover, we propose a
technique to extract the charge carrier density of state.

A specific spatial and energetic configuration of sites is
described by its density of states (DOS) and by its spatial
distribution function. Here we assume a homogeneous spatial
distribution and a Gaussian DOS \cite{r17},
\begin{equation} g\left(
\varepsilon \right) = \frac{1}{{2\pi \sqrt \sigma  }}\exp \left( {
- \frac{{\left( {\varepsilon  - \varepsilon _0 } \right)^2
}}{{\sqrt {2\sigma ^2 } }}} \right)
\end {equation}
where $\sigma $ is defined as the width of the DOS, $\varepsilon $
 is the energy, and $\varepsilon _0 $ is the center of the DOS. We use the Gaussian functional form to
enable comparison with other methods. However, we do not expect a
change in the principle results discussed here, while replacing
the Gaussian DOS with a more accurate (experimentally determined)
DOS. The second assumption we use is that the charge transfer
process drives the charge carrier population towards equilibrium
energy-distribution. To clearly define an equilibrium
energy-distribution we assume that each state cannot contain more
than one charge carrier (due to columbic repulsion). Therefore the
equilibrium distribution function is given by the following
Fermi-Dirac form:
    \begin{equation}
    f\left( {\varepsilon ,\eta } \right) = {1 \mathord{\left/
 {\vphantom {1 {\left[ {1 + \exp \left( {\beta \left( {\varepsilon _i  - \eta } \right)} \right)} \right]}}} \right.
 \kern-\nulldelimiterspace} {\left[ {1 + \exp \left( {\beta \left( {\varepsilon _i  - \eta } \right)} \right)} \right]}}
    \end{equation}
where $\eta $ is the chemical potential. A transfer rate which
will drive the charge carriers toward a detailed equilibrium
described in Eq. (2), has the general form of:
    \begin{equation}
\upsilon _{ij}  = \upsilon \left( {\left| {\varepsilon _j  -
\varepsilon _i } \right|,{\bf{R}}_{{\bf{ij}}} } \right)\exp \left[
{ - \frac{\beta }{2}\left[ {\left( {\varepsilon _j  - \varepsilon
_i } \right) + \left| {\varepsilon _j  - \varepsilon _i } \right|}
\right]} \right] \end{equation}
 where $\varepsilon _i,\varepsilon_j ,{\bf{R}}_{{\bf{ij}}} $
 are the initial energy, final energy, and the initial-final states vector, respectively; $\upsilon $
 is an envelope function that depends only on the absolute energy difference and the spatial coordinates, and $\beta $
 is the inverse temperature ($1/kT$). In the absence of an
additional energy dissipation process, and when the form of the
transfer rate deviates from Eq. (3) (as in the adiabatic polaronic
rate), the charge carriers will never reach thermal equilibrium
(Eq. (2)). On the other hand, when the envelope function is
decaying exponentially in space, with an inverse localization
radii $\gamma$ ($\upsilon  = \upsilon _0 \exp \left( { - \gamma
\left| {{\bf{R}}_{{\bf{ij}}} } \right|} \right)$), the transfer
rate becomes the known Miller-Abrahams rate \cite{r2}, and the
charge carriers reach equilibrium after a sufficient time has
lapsed.

In order to calculate the transport properties one needs to solve
the Master Equation determined by the transfer rate and the
energetic and spatial configuration of states, and to average over
all possible configurations. Instead, we propose to use the mean
medium approximation (MMA) where averaging on all possible
discrete configurations is replaced by solving the transport
properties of a homogeneous and continuous media, characterized by
the energetic DOS function. Such an approximation is valid only
for large enough space and time scales, and where the excess
energy gained by the accelerating field is dissipated fast. Under
the conditions for which the MMA is valid, one can deduce: 1) the
charge carrier population is near detailed equilibrium, and the
occupation probability of each site is determined by the local
quasi-chemical potential and the equilibrium distribution
function. 2) Since the medium is homogeneous all points in space
are characterized by the same charge concentration and transfer
rates and hence, the relaxation of a charge excess must follow the
Poissonian Debye pattern. Namely, the transport has to be Gaussian
(Markovian) and can be characterized by the first two spatial
moments of the transport Green function which are related to the
mobility ($\mu $) and the diffusion coefficient ($D$). Based on
detailed equilibrium and given a DOS and a charge concentration,
the Einstein relation ($D/\mu $) can be uniquely derived. As the
system is Markovian, one needs only to derive a value for the
mobility as a function of charge density and electric field to
complete the transport description. We perform this calculation
below using the MMA framework.

Before embarking on the mobility calculation we emphasize the
importance of charge concentration effects in real devices. We
calculated the charge carrier distribution in energy, $p\left(
\varepsilon \right)$, at equilibrium (Fig. 1) for several chemical
potentials ($\eta $). As long as the chemical potential is below
$\varepsilon _0  - \sigma \left( {\beta \sigma + 2} \right)$ the
charge concentration has a Gaussian distribution centered at
$\varepsilon _0  - \beta \sigma ^2$, and the Boltzmann
approximation is valid. For chemical potentials above this value
the charge carrier distribution deviates from the Gaussian shape
and the average energy increases. At this point the Boltzmann
approximation is no longer valid, and one can expect the transport
properties of the charge carriers to change. The relevant charge
concentration can be then calculated for a given DOS. The total
DOS can be estimated as the maximum concentration of electronic
units (sites) or one over the molecular volume of such unit (5-10
monomers or ~300 atoms volume equivalent to a total DOS of
$\sim$10$^{20}$ [cm$^{-3}$]) \cite{r19}. For a DOS width of 5$kT$
(130 meV at room temperature), and a chemical potential of
$\varepsilon _0 - \sigma \left( {\beta \sigma + 2} \right)$, the
corresponding charge concentration is approximately 10$^{11}$
[cm$^{-3}$]. As polymer based electronic devices (as LEDs and
FETs) operate at charge concentrations beyond 10$^{15}$
[cm$^{-3}$], the charge concentration effects should be taken into
account. It is evident that for any practical concentration the
charge energy distribution in a Gaussian DOS does not follow the
Boltzmann distribution function. Previously we have shown that the
$D/\mu $ ratio diverges from the classical $kT/q$ value
\cite{r20}. In the following we will consider the effect of the
charge concentration on the mobility.
\begin{figure}
\includegraphics[scale=0.4]{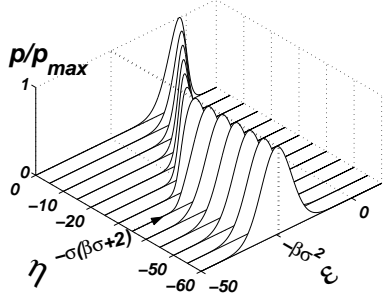}
\caption{\label{fig1} The normalized charge carrier energetic
distribution ($p\left( \varepsilon  \right)$) at various chemical
potentials ($\eta$), for a Gaussian DOS ($\sigma=5/\beta$)
centered at energy $\varepsilon$=0. For $\eta > \varepsilon _0 -
\sigma \left( {\beta \sigma  + 2} \right)$, the energetic charge
carrier distribution is modified when compared to the
non-degenerate distribution.}
\end{figure}

The total current is calculated by integrating the current between
each two sites while applying the MMA assumption that each point
contains all possible energy states with a relative weight
determined by the DOS function ($g\left( \varepsilon \right)$):
\begin{equation}
\begin{array}{l}
 {\bf{J}} = \int\limits_\Re  {d{\bf{R}}_{{\bf{ij}}} } \int\limits_{ - \infty }^\infty  {d\varepsilon _i } \int\limits_{ - \infty }^\infty  {d\varepsilon _j } \upsilon _{ij} \left( {{\bf{R}}_{{\bf{ij}}} ,\varepsilon _i ,\varepsilon _j } \right)g\left( {\varepsilon _i } \right)f\left( {\varepsilon _i ,\eta } \right)g\left( {\varepsilon _j  - {\bf{R}}_{{\bf{ij}}}  \cdot {\bf{E}}} \right) \\
 \left[ {1 - f\left( {\varepsilon _j  - {\bf{R}}_{{\bf{ij}}}  \cdot {\bf{E}},\eta } \right)} \right]{\bf{R}}_{{\bf{ij}}}  \cdot {\bf{\hat E}} \\
 \end{array}
\end{equation}
where ${\bf{R}}_{{\bf{ij}}}  \cdot {\bf{\hat E}}$ is the potential
drop between sites induced by the electric field E. Inserting the
Miller Abrahams hopping rate and selecting the $\left( {\left|
{{\bf{R}}_{{\bf{ij}}} } \right|,z,\varphi } \right)$
 coordinate system, where $z = {\bf{R}}_{{\bf{ij}}}  \cdot {\bf{\hat E}}$
 and $\varphi $
 is the angle in the plane perpendicular to z, the current integral becomes dependent on one spatial dimension, and reads:
 \begin{equation}
\begin{array}{l}
 {\bf{J}} = 2\pi \int\limits_{ - \infty }^\infty  {dz} \int\limits_{ - \infty }^\infty  {d\varepsilon _i } \int\limits_{ - \infty }^\infty  {d\varepsilon _j } g\left( {\varepsilon _i } \right)f\left( {\varepsilon _i ,\eta } \right)g\left( {\varepsilon _j  - \left( {zE} \right)} \right)\left[ {1 - f\left( {\varepsilon _j  - \left( {zE} \right),\eta } \right)} \right] \\
 \exp \left( { - \frac{\beta }{2}\left( {\left( {\varepsilon _j  - \varepsilon _i } \right) + \left| {\varepsilon _j  - \varepsilon _i } \right|} \right)} \right)z\exp \left( { - \gamma \left| z \right|} \right)\left( {\gamma \left| z \right| + 1} \right) \\
 \end{array}
\end{equation}
The mobility is calculated by the definition $\mu \equiv {{\bf{J}}
\mathord{\left/{\vphantom {{\bf{J}} {p{\bf{E}}}}} \right.
\kern-\nulldelimiterspace} {p{\bf{E}}}}$, where p is the total
charge carrier concentration: $p\left( \eta  \right) = \int
{g\left( \varepsilon  \right)f\left( {\varepsilon ,\eta }
\right)d\varepsilon }$ . The charge density dependence of the
mobility for a range of electrical fields is shown in Fig. 2. We
note that the effect of the charge density becomes pronounced for
$\eta  > \varepsilon _0 - \sigma \left( {\beta \sigma  + 2}
\right)$ where the system is degenerate, namely non-diluted (see
also Fig. 1). At very high charge concentration the mobility value
has a maximum due to "over filling" of the states to the point
where there are less states left to transfer into. The figure also
shows that at higher electric fields the density dependence is
less pronounced and the effect of the DOS form diminishes. At low
electric fields, the curve describing the mobility is determined
by the shape of the DOS. This can be shown by linearizing Eq. (5)
in the limit $\beta |zE|<<1$:
\begin{equation}
\mu _{E \to 0}  = \frac{{32\pi \beta \upsilon _0 }}{{q\gamma ^5
\cdot p\left( \eta ,g\left( \varepsilon  \right)  \right)
}}{\int\limits_{ - \infty }^\infty  {d\varepsilon _i
\int\limits_{\varepsilon _i }^\infty  {d\varepsilon _j g\left(
{\varepsilon _i } \right)f\left( {\varepsilon _i ,\eta }
\right)g\left( {\varepsilon _j } \right)\left[ {1 - f\left(
{\varepsilon _j ,\eta } \right)} \right]\exp \left( { - \beta
\left( {\varepsilon _j  - \varepsilon _i } \right)} \right)} } }
\end{equation}
The low field mobility, given in Eq. (6), depends only on the
exact shape of the DOS. Therefore, the DOS shape can be extracted
from measurements of the low field mobility dependence on charge
concentration, using this equation.

In Fig. 3 the mobility dependence on the electrical field is
illustrated, for several charge concentrations. It shows that at
the low concentration region all the mobility curves merge to one
curve which depends strongly on the field, as expected from the
Boltzmann approximation. However, as the charge concentration
increases (to a practical value) and the chemical potential
crosses the value of $\varepsilon _0  - \beta \sigma \left(
{\sigma + 2} \right)$, the low field mobility increases, and the
mobility field dependence weakens. At high electrical fields the
curves merge again, as the potential drop due to the applied field
shifts most of the final sites below that of the initial site
("state saturation").  In any case we expect that the theory used
here would not be strictly valid at the extreme cases of high
charge concentration and/or high electric fields. For example, it
was suggested that at the high field regime the polaron
dissociates into a free electron and a structural conformation
which is left behind \cite{r22}, hence a different hopping
mechanism should be applied. This can also be viewed as a
violation of our assumption that any excess energy gained by the
field is dissipated fast enough.
\begin{figure}
\includegraphics[scale=0.4]{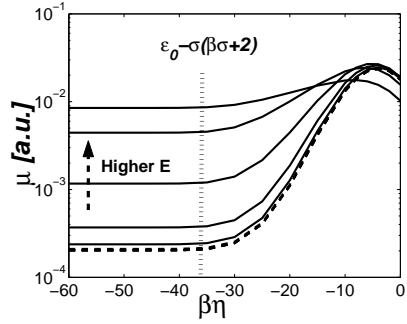}
\caption{\label{fig2} The mobility versus the chemical potential
(for DOS width  $\sigma=5/\beta$). The mobility does not change at
the diluted system region ($\eta < \varepsilon _0 - \sigma \left(
{\beta \sigma  + 2} \right)$), related to charge concentration of
$\sim$10$^{11}$ [cm$^{-3}$]).}
\end{figure}
\begin{figure}
\includegraphics[scale=0.4]{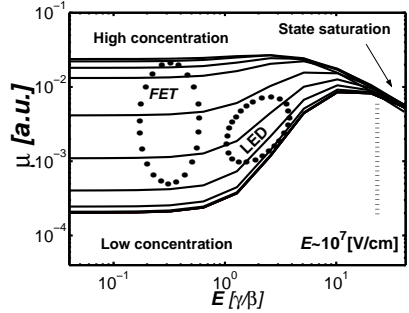}
\caption{\label{fig3} The mobility versus the electrical field for
different charge concentration (for DOS width $\sigma=5/\beta$).
The curves for low charge concentrations ($\eta < \varepsilon _0 -
\sigma \left( {\beta \sigma  + 2} \right)$) unite. At high charge
concentration the mobility remains high and almost independent of
the electrical field at most of the electrical field range.}
\end{figure}
The inconsistency between the measured mobility in LEDs and FETs
that was pointed out at the beginning of this paper can now be
explained using the MMA calculation results. We propose that it
steams from the vastly different operating conditions of each
device expressed as the typical charge concentration and electric
field. The ovals in Fig. 3 denote the typical operation field and
concentration ranges in LEDs and FETs. The low concentration, high
field condition in LEDs leads to a relatively low, strongly field
dependant mobility. On the other hand, in FETs the mobility is
typically higher and its dependence on the field is weak, as the
gate bias induces a high charge concentration (as was demonstrated
in \cite{r10}). Similarly, another difference between FETs and
LEDs is observed while examining the influence of charge carrier
concentration on the activation energy of the mobility (Fig. 4).
Since the transfer rate of the model is in the Miller-Abrahams
form the only possible origin of the activation energy is the
energetic disorder (contrary to other possible contribution as the
polaronic effect, e.g. Ref. \cite{r23}). While the activation
energy at high applied fields is low, at low applied fields it is
strongly dependent on the charge carrier concentration (insert in
Fig. 4). The difference of ~200 meV between "low concentration"
(LED) mobility activation energy and "high concentration" (FET)
activation energy was measured by us (215$\pm$10 meV \cite{r24}
and 400$\pm$40 meV \cite{r11} in MEH-PPV based FET and LED,
respectively) and in Ref. \cite{r12}. However, the activation
energy measured at high concentrations (FET) is significantly
higher than the GDM, Miller-Abrahams model prediction, indicating
an additional source for the activation energy besides the
energetic disorder. Explicitly, the activation energy in the
polaronic model is constituted from the energetic disorder
contribution and half of the polaronic energy. Therefore, at high
concentration measurement, where the energetic disorder
contribution is negligible, the polaronic binding energy can be
estimated as twice the measured activation energy ($\sim$400 meV
for MEH-PPV at the previous example).
\begin{figure}
\includegraphics[scale=0.4]{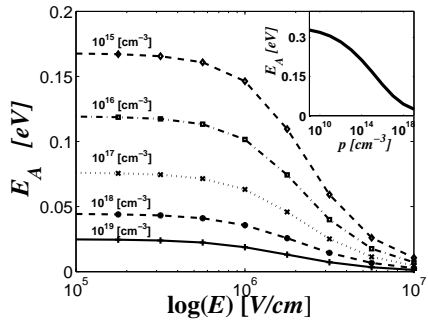}
\caption{\label{fig4} The activation energy of the mobility versus
the electrical field for different charge concentration (for DOS
width $\sigma$=130 meV, $N_V$=10$^{20}$ [cm$^{-3}$] and Miller
Abrahams transfer between sites). The activation energy at low
electrical field (plotted in the insert) is strongly dependent on
the charge concentration and diminishes at high charge
concentration.}
\end{figure}

In conclusion, we present a unified calculation method for
variable range hopping of diluted and non-diluted materials (low
and high charge carrier concentration), and a varying applied
electric field. We demonstrate that the major differences between
the transport properties measured at high concentration (in FETs)
and low concentration (in LEDs) can be explained by this model, in
particular the difference in the mobility and the activation
energy of the mobility between the two. Moreover, the low value of
the predicted high concentration activation energy, where compared
to measurements, enables us to estimate the polaron binding energy
in MEH-PPV as 400$\pm$40 meV. A theoretical method to extract the
charge carrier DOS from the charge transport measurements is
proposed, as well. Finally, we note that charge concentration
effects play a role also in other experimental situations such as
time of flight experiments \cite{r25}.

This research (No. 56/00-11.6) was supported by the Israel science
foundation. Y.R. thanks Israel science Ministry and Israel science
foundation for the generous scholarships.


\begin{thebibliography}{10}
\bibitem{r17}
For simplicity we do not introduce here the notion of spatial
distribution. however, for most of the results presented here it
will only scale down the absolute electric field values quoted in
the text.
\end{thebibliography}
\end{document}